\documentclass[nofootinbib,aps,amssymb,floatfix,twocolumn,superscriptaddress,preprintnumbers]{revtex4}

\usepackage{graphicx}
\usepackage{bm}
\usepackage{amsmath}
\usepackage{amssymb}
\usepackage{amsfonts}
\usepackage{float}
\usepackage{hyperref}
\usepackage{comment}
\usepackage{dsfont}  
\usepackage{slashed}  
\usepackage{booktabs}
\usepackage{multirow}
\usepackage{subfigure}
\usepackage[sort&compress]{natbib}
\usepackage{xcolor}
\usepackage{ulem}

\newcommand{\be}{\begin{equation}}  
\newcommand{\ee}{\end{equation}}  
\newcommand{\beq}{\begin{eqnarray}} 
\newcommand{\eeq}{\end{eqnarray}}

\newcommand{\bea}{\begin{eqnarray}}
\newcommand{\eea}{\end{eqnarray}}

\begin{document}

\title{Quark Reggeization in QCD from the Wilson line formalism}

\author{Tolga Altinoluk}
\affiliation{ National Centre for Nuclear Research, Pasteura 7, Warsaw 02-093, Poland}

\author{Guillaume Beuf}
\affiliation{ National Centre for Nuclear Research, Pasteura 7, Warsaw 02-093, Poland}

\author{Jules Favrel}
\affiliation{ National Centre for Nuclear Research, Pasteura 7, Warsaw 02-093, Poland}

\author{Michael Fucilla}
\affiliation{ National Centre for Nuclear Research, Pasteura 7, Warsaw 02-093, Poland}

\begin{abstract}
We establish a Wilson-line formulation of quark Reggeization in QCD. An interpolating operator for the Reggeized quark is identified in terms of semi-infinite Wilson lines, and its nonlinear rapidity evolution is derived. We provide strong evidence that this operator remains an eigenstate of the rapidity evolution up to next-to-leading logarithmic accuracy. Within the leading logarithmic accuracy, we explicitly recover the one-loop quark Regge trajectory. Our results provide a first-principles operator framework for quark Reggeization and pave the way for a systematic all-order description of high-energy QCD amplitudes with $t$-channel quark exchange in terms of Wilson lines.
\end{abstract}

\maketitle

{\it Introduction}---The high-energy limit of QCD exhibits an emergent structure in which the asymptotic energy dependence of scattering amplitudes is encoded in Reggeized quark and gluon exchange. This paradigm is most prominently realized by gluon Reggeization~\cite{Lipatov:1976zz} and the associated Balitsky-Fadin-Kuraev-Lipatov (BFKL) equation~\cite{Fadin:1975cb,Kuraev:1976ge,Kuraev:1977fs,Balitsky:1978ic}, which underpin much of our understanding of high-energy scattering and small-$x$ dynamics. In the Wilson-line formulation of high-energy QCD, gluon Reggeization acquires a transparent operator interpretation in terms of eikonal Wilson lines and their rapidity evolution~\cite{Caron-Huot:2013fea}. By contrast, the Wilson-line operator underlying quark Reggeization has remained elusive, preventing a first-principles derivation of its rapidity evolution within this framework.

In the high-energy limit $s\simeq -u \gg |t|$, scattering amplitudes with fixed $t$-channel quark or gluon quantum numbers exchange acquire a Regge-pole form. For color-octet amplitudes of negative signature, one has~\cite{Lipatov:1976zz}
\begin{equation}
{\cal A}_{AB \rightarrow A'B'}^{(8,-)} =
\Gamma_{A} \,  \frac{s}{t}
\left[
\left( \frac{s}{-t} \right)^{\omega (t)}
+
\left( \frac{-s}{-t} \right)^{\omega (t)}
\right]
\Gamma_{B} \; ,
\label{Pole:Eq:GluonReggeForm}
\end{equation}
where $1+\omega(t)$ denotes the gluon Regge trajectory. At one loop and in dimension $D= 2+d =4-2\epsilon$,
\begin{equation}
\omega^{(1)}(t)
= - \frac{g^2 N_c \Gamma(1+\epsilon)}{(4\pi)^{2-\epsilon}} \left(\frac{\boldsymbol{p}^{2}}{\mu^2}\right)^{-\epsilon}
\frac{\Gamma^2(-\epsilon)}{\Gamma(-2\epsilon)} \; ,
\label{GluonReggeTraj}
\end{equation}
with $ \boldsymbol{p}^{2} \simeq - t \; $, and $\Gamma_A$ and $\Gamma_B$ are energy-independent effective vertices describing the coupling of the external particles to the exchanged Reggeized gluon. This Regge form has been established to all orders at leading (LLA) and next-to-leading (NLLA) logarithmic accuracy~\cite{Fadin:1995xg,Fadin:1996tb,Braun:1999uz,Fadin:2000ww,Fadin:2002hz}. 

A similar phenomenon occurs for amplitudes carrying quark quantum numbers in the $t$-channel (color triplet and positive signature), which read~\cite{Fadin:1976nw,Fadin:1977jr}
\begin{equation}
\mathcal{A}_{A B \rightarrow A'B'}^{(3,+)}
=
\Gamma_{A} \;
\frac{\sqrt{s}}{-\slashed{q}_{\perp}}
\frac{1}{2}
\left[
\left(\frac{-s}{-t}\right)^{\delta(t)}
+
\left(\frac{s}{-t}\right)^{\delta(t)}
\right]
\Gamma_{B} \; ,
\end{equation}
where $1/2 + \delta(t)$ is the massless quark Regge trajectory and, at one-loop, reads
\begin{equation}
\delta^{(1)} (t) = \frac{C_F}{N_c}\,\omega^{(1)}(t)\;.
\label{Eq:QuarkReggeTraj}
\end{equation}
Quark Reggeization was rigorously proven at LLA~\cite{Bogdan:2006af} and was found to be compatible with the Regge form at NLLA through an explicit two-loop computation~\cite{Bogdan:2002sr}. More fundamentally, amplitudes mediated by Reggeized quark exchange are suppressed by one power of $\sqrt{s}$ relative to their gluon counterparts and therefore belong to the subeikonal sector of high-energy scattering. Since Wilson-line descriptions are naturally formulated at eikonal accuracy, the identification of a Wilson-line operator associated with the Reggeized quark requires extending the formalism into the subeikonal sector.

While Reggeization and the BFKL formalism successfully describe high-energy scattering in the dilute regime, the rapid growth of the amplitudes with energy requires nonlinear dynamics at sufficiently small Bjorken-$x$~\cite{Gribov:1983ivg}. In this regime, high-energy scattering is naturally formulated in terms of Wilson lines propagating through the strong color background (the QCD shockwave)~\cite{McLerran:1994vd,Balitsky:1995ub}, whose rapidity evolution is governed by the Balitsky–Kovchegov/Jalilian-Marian-Iancu-McLerran-Wiegert-Leonidov-Kovner (BK/JIMWLK) equations derived in \cite{Balitsky:1995ub,Kovchegov:1999yj,Kovchegov:1999ua,Jalilian-Marian:1996mkd,Jalilian-Marian:1997qno,Jalilian-Marian:1997jhx,Jalilian-Marian:1997ubg,Kovner:2000pt,Weigert:2000gi,Iancu:2000hn,Iancu:2001ad,Ferreiro:2001qy}. A breakthrough was achieved in ref.~\cite{Caron-Huot:2013fea}, which showed that gluon Reggeization at LLA and NLLA emerge naturally from the rapidity evolution of Wilson lines upon linearization. Whether an analogous correspondence exists for Reggeized quark exchange has remained an open question. 

Motivated by this correspondence, we investigate whether quark Reggeization admits an analogous Wilson-line description.  Since Reggeized quark exchange is intrinsically subeikonal, our construction relies on the recent extensions of shockwave formalism beyond the eikonal approximation~\cite{Altinoluk:2020oyd, Altinoluk:2024dba,Cougoulic:2022gbk,Borden:2024bxa,Balitsky:2015qba,Balitsky:2016dgz,Chirilli:2021lif,Chirilli:2026pkv,Chirilli:2026vij,Jalilian-Marian:2018iui,Jalilian-Marian:2017ttv,Boussarie:2021wkn,Boussarie:2023xun}. In this Letter, we identify an operator built from semi-infinite Wilson lines that carries quark quantum numbers in the $t$-channel and derive its nonlinear rapidity evolution using the background field method. In the dilute limit, we extract the linearized degree of freedom evolving according to the one-loop quark Regge trajectory. Projecting onto a positive signature operator, we identify a Reggeized quark interpolating operator, and we suggest that this latter Reggeizes up to next-to-leading logarithmic accuracy (NLLA). In the large-$N_c$ limit, quark Reggeization is observed without the need to separate positive and negative signatures, recovering the expected signature degeneracy. 

Our work fits into the broader program of understanding the all-order
structure of high-energy QCD amplitudes, which has so far predominantly
focused on gluon exchange (see~\cite{Lipatov:1991nf,Kirschner:1994gd,
Kirschner:1994xi,Chachamis:2013hma,Lipatov:2000se,Hentschinski:2011xg,Nefedov:2017qzc,Nefedov:2019mrg,
Rothstein:2016bsq,Moult:2022lfy,Gao:2024fyz,Gao:2024qsg,Moult:2017xpp,
Moult:2019vou} for alternative approaches). Beyond NLLA, Regge-pole
factorization breaks down through Regge cuts generated by multi-Reggeon
exchange~\cite{DelDuca:2001gu,DelDuca:2013ara,DelDuca:2014cya,
DelDuca:2017twk,Caron-Huot:2017fxr,Falcioni:2020lvv,DelDuca:2021vjq,
Caola:2021izf,Falcioni:2021buo,Falcioni:2021dgr,Byrne:2022wzk,
Fadin:2023roz,Buccioni:2024gzo,Abreu:2024xoh}. The Wilson-line formulation
provides a systematic framework for investigating these structures
~\cite{Caron-Huot:2013fea,Caron-Huot:2017fxr,Caron-Huot:2017zfo,
Caron-Huot:2020grv,Falcioni:2020lvv,Falcioni:2021buo,Falcioni:2021dgr};
see also Refs.~\cite{Fadin:2017hnr,Fadin:2017nka,Fadin:2023aen} for the
complementary diagrammatic approach. The present work extends this program
to $t$-channel quark exchange. \\

{\it The Shockwave formalism}---We formulate high-energy scattering in a light-cone basis defined by two null vectors $n_1$ and $n_2$, satisfying $n_1\!\cdot\! n_2=1$, and decompose four-vectors according to Sudakov variables as $k^{\mu} = k^+ n_1^\mu + k^- n_2^{\mu} + k_{\perp}^{\mu}$.
Note that transverse momenta in Euclidean space are denoted in bold characters, while Minkowskian transverse vectors carry the $\perp$ subscript. We consider the Regge limit $s\gg |t| $ and work in a frame where a projectile particle moves along the $+$ direction while the target system carries a large momentum component along the $-$ direction. We adopt the light-cone gauge condition $A\!\cdot\!n_2=0$.

In the shockwave approach~\cite{McLerran:1994vd,Balitsky:1995ub}, the gluon field is separated into external background modes, $\mathcal{A}^{\mu} (k)$, and internal quantum fluctuations, $A^{\mu} (k)$, according to their longitudinal momentum. More precisely, modes with $k^+<e^\eta p_{p}^+$ are included in the background field, while modes above this arbitrary rapidity cutoff are treated perturbatively, with $p_{p}^+ \sim \sqrt{s}$ and $\eta<0$. After boosting the target from its rest frame into the projectile frame, Lorentz contraction localizes the target background field into a thin sheet in light-cone time,
\begin{equation}
 \mathcal{A}^\mu(z)
=
\mathcal{A}^-(z^+,z^-=0,z_\perp)\,
n_2^\mu
\sim
\delta(z^+)\,
\mathcal{A}^{-}(\boldsymbol{z})
\,n_2^\mu ,
\end{equation}
which defines the so-called shockwave approximation.

The interaction of energetic partons with this background resums into Wilson lines extending along the projectile trajectory,
\begin{equation}
\hspace{-0.2 cm}\mathcal{U}_R(z_1^+,z_2^+,\boldsymbol{z})
=
\mathcal{P}
\exp
\left(
ig
\int_{z_2^+}^{z_1^+}
dz^+\,
    \mathcal{A}^{-a}(z^+,\boldsymbol{z})
T_R^a
\right) 
\label{Eq:GenericWilsonLine}
\end{equation}
where $\mathcal{P}$ denotes the path ordering and $T_R^a$ is the color generator in representation $R$. In the following, $\mathcal{U}_F$ and $\mathcal{U}_A$ denote Wilson lines in the fundamental and adjoint representations, respectively, and we use the compact notation $\mathcal{U}_R(\boldsymbol{z})$ to denote any Wilson line (or operator constructed from them), whose support is from $-\infty^+$ to $\infty^+$. Scattering is described in terms of matrix elements of Wilson-line operators, while their energy dependence is generated by the renormalization-group evolution in the rapidity cutoff $\eta$. \\

\begin{figure}
    \centering
    \includegraphics[width=0.4\linewidth]{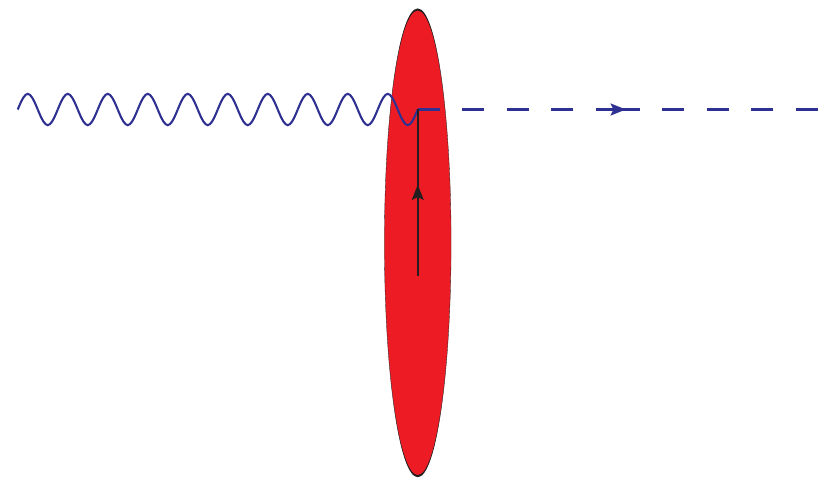}
    \hspace{1.4 cm}\includegraphics[width=0.4\linewidth]{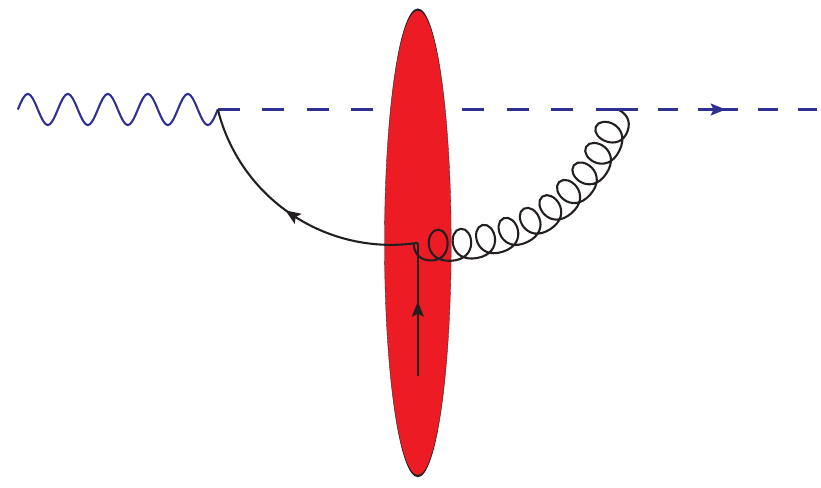} \\
    (a) \hspace{4.4 cm} (b)
    \caption{(a) A fast-moving photon interacting with the target quark background field and converting into a fast-moving quark, which is subsequently dressed by eikonal gluon interactions encoded in the shockwave. (b) A diagram contributing to the one-loop corrections to the process.}
    \label{fig:QIOLeading}
\end{figure}

Quark exchange in the $t$-channel is naturally suppressed in the high-energy limit, thus representing a particular class of subeikonal effects to the shockwave formalism~\cite{Balitsky:2015qba,Kovchegov:2016zex,Chirilli:2018kkw,Altinoluk:2025ang,Altinoluk:2023qfr}. In analogy to the gluon case, the quark field must be separated into fast, $\psi ( k^+ > e^\eta p_{p}^+, k^-, k_{\perp})$, and slow, $\Psi ( k^+ < e^\eta p_{p}^+, k^-, k_{\perp}) $, modes according to the rapidity cutoff. 
Processes involving quark exchange in the $t$-channel are described by operator structures that extend the standard eikonal Wilson line, which are identity-preserving operators. In particular, transitions in which an external probe changes its identity through $t$-channel quark exchange (see e.g. fig.~\ref{fig:QIOLeading}) are naturally encoded by operators built from semi-infinite Wilson lines and the target quark field, $\mathcal{O}[\mathcal{U}_R, \Psi]$ .\\

{\it The Reggeized gluon from eikonal Wilson lines}---Before constructing the Reggeized quark operator, it is useful to recall how the Reggeized gluon emerges from the Wilson line formalism. The Reggeized gluon interpolating (RGI) operator is defined as the logarithm of an infinite eikonal Wilson line in the adjoint representation~\cite{Caron-Huot:2013fea}. While eikonal scattering amplitudes are naturally formulated in terms of infinite Wilson lines, for later convenience we introduce the slightly generalized operator
\begin{gather}
 R^a (z_1^+, z_2^+, \boldsymbol{z}) \equiv \frac{f^{abc}}{g N_c} [\ln \mathcal{U}_A (z_1^+, z_2^+, \boldsymbol{z})]^{bc} \; .
 \label{Eq:ReggeizedGluon}
\end{gather}
The operator $R^a (z_1^+, z_2^+, \boldsymbol{z})$ starts at order $g^0$ in perturbation theory and interpolates a single free gluon. Beyond LO, it generates an ordered cascade of gluon emissions in the light-cone time $z^+$ (see eq.~(2.8) of ref.~\cite{Caron-Huot:2013fea}). The operator satisfies two key properties: 
\begin{itemize}
    \item[i.] A Wilson line in an arbitrary representation admits the expansion
    \begin{gather}
        \mathcal{U}_R (z_1^+, z_2^+, \boldsymbol{z} ) = e^{i g T^a_R R^a (z_1^+, z_2^+, \boldsymbol{z} )}  \nonumber \\  = 1 + ig T^a_R R^a (z_1^+, z_2^+, \boldsymbol{z} ) + \mathcal{O} (g^2 R^2) \; .
    \end{gather}
    In the dense regime of QCD, $g R \sim 1$, the full nonlinear Wilson-line structure must be retained. By contrast, in the dilute regime $gR \ll 1$, the theory admits an expansion in terms of Reggeized gluon degrees of freedom.  
    \item[ii.] It carries a negative signature, $\displaystyle R^{a} (z_2^+, z_1^+, \boldsymbol{z}) = - R^{a} (z_1^+, z_2^+, \boldsymbol{z})$ which, in the Regge limit, corresponds to antisymmetry of the scattering amplitudes in eq.~(\ref{Pole:Eq:GluonReggeForm}) under the exchange $s \leftrightarrow u \approx -s$. In the Wilson-line language, this property reflects antisymmetry under the interchange of initial and final states, corresponding to the exchange of the $x^+$-endpoints in the operator. Indeed, reversing the path in eq.~(\ref{Eq:ReggeizedGluon}) corresponds to the transformation $\mathcal{U}_A^{bc} (z_1^+, z_2^+, \boldsymbol{z}) \rightarrow \mathcal{U}_A^{cb} (z_1^+, z_2^+, \boldsymbol{z}) = [\mathcal{U}_A^{bc} (z_1^+, z_2^+, \boldsymbol{z})]^{\dagger} = \mathcal{U}_A^{bc} (z_2^+, z_1^+, \boldsymbol{z})$.
\end{itemize}

The defining property of $R^a$ is its rapidity renormalization-group evolution in the cutoff $\eta$, which follows directly from the evolution equation of a single Wilson line. The latter has been known for a long time at both leading (LO)~\cite{Balitsky:1995ub} and next-to-leading (NLO)~\cite{Balitsky:2013fea} accuracy. Adopting dimensional regularization, at LO and in the fundamental representation, one finds
\begin{gather}
    \frac{\partial}{ \partial \eta } [\mathcal{U}_{F} (\boldsymbol{z}_1)]_{ij} =  a_s  \int d^d \boldsymbol{z}_{2} \frac{{\rm Tr_c} [\mathcal{U}_{F} (\boldsymbol{z}_1) \mathcal{U}_{F}^{\dagger} (\boldsymbol{z}_2) ]}{(\boldsymbol{z}_{12}^2)^{d-1}} [\mathcal{U}_{F} (\boldsymbol{z}_2)]_{ij}  ,
    \label{Eq:SingleWilsonLineEvo}
\end{gather}
where we neglected scaleless integrals through the choice $\epsilon_{\rm UV} = \epsilon_{\rm IR} \equiv \epsilon$ and we introduced the notations $\boldsymbol{z}_{ij} \equiv \boldsymbol{z}_i-\boldsymbol{z}_j$ and
\begin{gather*}
    a_s = \frac{g^2 (\mu^2)^{1-d/2}}{8 \pi^{1+d}} \left[\Gamma \left( \frac{d}{2} \right) \right]^2
\end{gather*}
In the dilute limit, $gR \ll 1$, the evolution can be linearized through eq.~(\ref{Eq:ReggeizedGluon}) and projected onto the one-Reggeized-gluon sector, yielding
\begin{gather}
    \frac{\partial}{ \partial \eta } R^a (\boldsymbol{z}_1) =  N_c \; \mathcal{K}_{\rm RP} (\boldsymbol{z}_{1}, \boldsymbol{z}_{2}) \otimes_{\boldsymbol{z}_2} R^a (\boldsymbol{z}_2) +
\mathcal{O}(g^4 R^3) \; ,
    \label{Eq:ReggeGluonEvoCoord}
\end{gather}
where
\begin{gather}
    \mathcal{K}_{\rm RP} (\boldsymbol{z}_{1}, \boldsymbol{z}_{2}) =  \frac{a_s  }{(\boldsymbol{z}_{12}^2)^{d-1}} \; ,
    \label{Eq:KRegge}
\end{gather}
is the Regge-pole (RP) coordinate space kernel.
Here, we introduced the compact notation $f(\boldsymbol{z}) \otimes_{\boldsymbol{z}} g(\boldsymbol{z}) = \int d^d \boldsymbol{z} \; f(\boldsymbol{z}) g (\boldsymbol{z})$.
Passing to momentum space diagonalizes the evolution kernel and leads to
\begin{gather}
    \frac{\partial}{ \partial \eta } R^a (\boldsymbol{p}) =  \omega^{(1)} (-\boldsymbol{p}^2) R^a (\boldsymbol{p}) +
\mathcal{O}(g^4 R^3)
\end{gather}
which is governed by the one-loop gluon Regge trajectory defined in eq.~(\ref{GluonReggeTraj}). Gluon Reggeization thus acquires a natural interpretation within the Wilson-line formalism: the Regge trajectory emerges as the eigenvalue governing the linearized rapidity evolution of eikonal degrees of freedom. \\

{\it Towards the Reggeized quark from semi-infinite Wilson lines}---Unlike gluon exchange,  the $t$-channel quark exchange is intrinsically subeikonal in nature. It is suppressed by a factor of $\sqrt{s}$ relative to gluon exchange and cannot be described solely in terms of infinite eikonal Wilson lines.
To identify the dominant high-energy contribution, we decompose the target quark background field, $\Psi$, into its good, $\Psi^{(-)}\equiv (\gamma^+\gamma^-/2) \Psi$, and bad, $\Psi^{(+)}\equiv (\gamma^-\gamma^+/2) \Psi$, light-cone components. Under a longitudinal boost by a factor $\gamma_t$ along the $x^-$ direction, these fields scale as $\Psi^{(-)}\sim\gamma_t^{1/2}$ and $\Psi^{(+)}\sim\gamma_t^{-1/2}$, respectively.  Therefore, the leading contribution to quark exchange in the Regge limit is entirely controlled by operators built from $\Psi^{(-)} (z)$. Moreover, as in the gluon case, at leading eikonal accuracy, the target quark field becomes localized into a shockwave configuration,
\begin{gather}
\Psi^{(-)}(z) =
\Psi^{(-)}(z^+, z^-= 0, z_{\perp})
\sim
\delta(z^+) \Psi^{(-)} (\boldsymbol{z}) \; ,
\end{gather}
reflecting Lorentz contraction along the longitudinal direction.

Guided by the preceding discussion, we seek the simplest Wilson-line operator interpolating an outgoing quark emerging from the target and dressed by an arbitrary number of soft gluon interactions. These requirements motivate the process $\gamma q\rightarrow  q$ illustrated in the left diagram of fig.~\ref{fig:QIOLeading}, leading to the operator
\begin{gather}
Q_i(\boldsymbol{z})
=
\int
dz^+\,
[\mathcal{U}_F(\infty^+,z^+,\boldsymbol{z})]_{ij}
\,
\Psi_j^{(-)}(z^+,\boldsymbol{z}) \; .
\label{Eq:AlmostReggeizedQuarkInter}
\end{gather}

The operator consists of a semi-infinite Wilson line in the fundamental representation starting at the insertion point of the good component of the quark field. The Wilson line resums multiple eikonal interactions of the fast-moving quark with the target gluon background, while the insertion of $\Psi^{(-)} (z)$ generates the conversion of the projectile into a state carrying $t$-channel quark quantum numbers. 

Linearizing eq.~(\ref{Eq:AlmostReggeizedQuarkInter}) in RGI operators gives
\begin{gather}
Q_i(\boldsymbol{z})
=
\int
dz^+
\left[
\delta_{ij}
+
ig\,t^a_{ij}
R^a(\infty^+,z^+,\boldsymbol{z})
\right]
\Psi_j^{(-)}(z^+,\boldsymbol{z})
\nonumber
\\ +
\mathcal{O}(g^2R^2)
\equiv
R_{Q,i}(\boldsymbol{z})
+
\mathcal{O}(g^2R^2)\; .
\label{Eq:AlmostReggeizedQuarkInterLinear}
\end{gather}
The operator $R_{Q,i}$, therefore, plays for the quark exchange the same role that the RGI operator plays for gluon exchange. At LO, it interpolates a free quark, while its higher-order terms are described by its coupling to the Reggeized gluon background. It thus provides a natural interpolating-operator candidate for a $t$-channel quark exchange in the dilute regime.

As we shall show, recovering genuine Regge-pole behaviour requires a projection onto operators of definite signature. To this end, we introduce a signature transformation $\mathcal{S}$, defined as the $\infty^+ \leftrightarrow -\infty^+$ exchange in the operators, and corresponding to the $s\leftrightarrow u$-channel crossing. While the Reggeized gluon interpolating operator in eq.~(\ref{Eq:ReggeizedGluon}) is an eigenstate of $\mathcal{S}$ with eigenvalue $-1$, the operator obtained from eq.~(\ref{Eq:AlmostReggeizedQuarkInterLinear}) is not. Indeed, acting with $\mathcal{S}$ on its Wilson-line realization, $Q_i(\boldsymbol{z})$, maps it into
\begin{gather}
\bar{Q}_i (\boldsymbol{z})
=
\int dz^+\,
[\mathcal{U}_F^{\dagger}(z^+,-\infty^+,\boldsymbol{z})]_{ij}
\,\Psi_j^{(-)}(z^+,\boldsymbol{z})
\nonumber\\
= \int dz^+\,\left[
\delta_{ij}
+
ig\,t^a_{ij}
R^a(-\infty^+,z^+,\boldsymbol{z})
\right] \Psi_j^{(-)}(z^+,\boldsymbol{z})  \nonumber \\ 
+ \mathcal{O}(g^2 R^2)  \equiv
\bar{R}_{Q,i}(\boldsymbol{z})
+ \mathcal{O}(g^2 R^2)\; .
\label{SignatureOperationQuark}
\end{gather}
This operator corresponds to the crossed-channel configuration $\bar{q} q\rightarrow \gamma $. \\

{\it Evolution equation of $Q_i(\boldsymbol{z})$ and Reggeization in the large-$N_c$ limit}---The rapidity evolution of the operator in eq.~(\ref{Eq:AlmostReggeizedQuarkInter}) can be derived within the background-field formalism. At one loop, eikonal power counting allows only one interaction vertex inside the medium (see e.g. Appendix C of ref.~\cite{Altinoluk:2025ivn}), leaving a single contributing diagram (see (b) in fig.~\ref{fig:QIOLeading}).

The derivation of the evolution equation of the operator $Q_i(\boldsymbol{z}_1)$ is presented in the accompanying paper~\cite{Altinoluk:2026}. The resulting equation reads 

\begin{gather}
    \frac{\partial Q_i (\boldsymbol{z}_1)}{ \partial \eta}
    = a_s
    \int d^d \boldsymbol{z}_2
    \frac{1}{(\boldsymbol{z}_{12}^2)^{d-1}}
    \frac{1}{2}
    \bigg\{
    \left(
    {\rm Tr}[
    \mathcal{U}_F (\boldsymbol{z}_1 )
    \mathcal{U}_F^{\dagger} (\boldsymbol{z}_2 )]
    \right. \nonumber \\ \left. -
    \frac{1}{N_c}
    \right) \delta_{ij}
    -  
    \frac{1}{N_c}
    \bigg(
    [
    \mathcal{U}_F (\boldsymbol{z}_1 )
    \mathcal{U}_F^{\dagger} (\boldsymbol{z}_2 )
    ]_{ij}
    -
    \delta_{ij}
    \bigg)
    \bigg\}
    Q_j (\boldsymbol{z}_2)\; .
\label{Eq:AlmostReggeEvoCoord}
\end{gather}

The relation between eq.~(\ref{Eq:AlmostReggeEvoCoord}) and quark Reggeization is not immediately apparent at finite $N_c$. We therefore begin by considering the planar limit.

As a one-loop consistency check, setting all Wilson lines to the identity reproduces, after Fourier transform, the evolution of the good quark component $\Psi^{(-)}(z)$, governed by the one-loop quark trajectory.

At all orders, the analysis greatly simplifies in the planar limit. The non-planar contribution to eq.~(\ref{Eq:AlmostReggeEvoCoord}) is suppressed and, upon restricting to the one-Reggeized-gluon sector, the dipole operator reduces to $N_c$. Equation~(\ref{Eq:AlmostReggeEvoCoord}) therefore immediately yields a Regge-pole evolution for $R_{Q,i}(\boldsymbol{z})$, governed by the large-$N_c$ quark trajectory, $\omega^{(1)}(-\boldsymbol{p}^2)/2$ (see eq.~(\ref{Eq:QuarkReggeTraj})), and valid up to $\mathcal{O} (g^4 R^2)$.

This simplification admits a natural interpretation in terms of signature. At finite $N_c$, and starting at the NLLA, fermion-exchange amplitudes develop a more intricate singularity structure than a simple Regge pole, with Reggeization emerging only upon projection onto states of definite positive signature~\cite{Fadin:1976nw,Fadin:1977jr,Bogdan:2002sr}. In the planar limit, the non-planar contributions responsible for this structure are suppressed, and Regge-pole behavior survives already at the level of the non-signatured amplitudes. Equation~(\ref{Eq:AlmostReggeEvoCoord}) provides a dipole-like realization of this phenomenon (see the accompanying paper~\cite{Altinoluk:2026} for further details). \\

{\it Signaturization and Reggeization at finite $N_c$}---We now return to the finite-$N_c$ case, where the operator $R_{Q,i}$ cannot be identified as the Reggeized quark interpolating (RQI), since it does not carry a definite (positive) signature. Indeed, beyond the large-$N_c$ limit, the second term in the curly brackets contributes already at $\mathcal{O}(g^3 R)$ in the dilute limit, leading to
\begin{gather}
    \frac{\partial R_{Q,i} (\boldsymbol{z}_1)}{ \partial \eta} = C_F \mathcal{K}_{\rm RP} (\boldsymbol{z}_{1}, \boldsymbol{z}_{2}) \otimes_{\boldsymbol{z}_2} R_{Q,i} (\boldsymbol{z}_2) \nonumber \\ + \frac{ \displaystyle [\mathcal{K}_{\rm RPB} (\boldsymbol{z}_{1}, \boldsymbol{z}_{2}) ]_{ij} }{2N_c}  \;\otimes_{\boldsymbol{z}_2} \big (R^a(\boldsymbol{z}_2) - R^a(\boldsymbol{z}_1) \big ) \bar{R}_{Q,j} (\boldsymbol{z}_2) ,
    \label{Eq:ReggePlusReggeBreak1}
\end{gather}
up to $\mathcal{O}(g^4 R^2)$ and where $ \displaystyle [\mathcal{K}_{\rm RPB} (\boldsymbol{z}_{1}, \boldsymbol{z}_{2}) ]_{ij} \equiv igt^a_{ij} \mathcal{K}_{\rm RP} (\boldsymbol{z}_{1}, \boldsymbol{z}_{2})$
represents the Regge-pole-breaking (RPB) part of the evolution. The RPB term starts to contribute at the NLLA and prevents the Regge-pole behavior of $R_{Q,i} (\boldsymbol{z})$~\cite{Altinoluk:2026}.

The structure of eq.~(\ref{Eq:ReggePlusReggeBreak1}) strongly suggests that the violation of Regge-pole behaviour is tied to the absence of definite signature. First, the evolution couples the operator $R_{Q,i}$ directly to its $s\leftrightarrow u$ crossed counterpart $\bar{R}_{Q,i}$. More importantly, eq.~(\ref{Eq:ReggePlusReggeBreak1}) possesses a simple transformation property under the signature operation $\mathcal{S}$: it exchanges $R_{Q,i}\leftrightarrow \bar{R}_{Q,i}$, leaves $\mathcal{K}_{\rm RP}$ and $\mathcal{K}_{\rm RPB}$ invariant, but changes the sign of the term in the second line of Eq.~(\ref{Eq:ReggePlusReggeBreak1}) as consequence of the negative signature of the RGI operator, $R^a (\boldsymbol{z})$. This immediately suggests that Regge-pole behaviour can be restored by projecting onto the positive signature operator. Following the standard construction for scattering amplitudes~\cite{Fadin:1977jr}, we thus construct the positive combination $ \displaystyle R_{Q,i}^{(+)} (\boldsymbol{z}) =  \left( R_{Q,i} (\boldsymbol{z}) + \bar{R}_{Q,i} (\boldsymbol{z}) \right) /2$. Neglecting corrections beyond the one-Reggeized-gluon sector, the sum and difference of the two RGI operators can be related to the logarithm of the product of two semi-infinite Wilson lines, through the Baker–Campbell–Hausdorff formula, yielding
\begin{widetext} 
\begin{gather}
    R_{Q,i}^{(+)} (\boldsymbol{z}) =   \int d z^+ 
    \, \left \{ \delta_{ij} + \frac{i f^{abc} t^a_{ij}}{2N_c} \bigg[ \ln \left( \mathcal{U}_A (\infty^+, z^+, \boldsymbol{z}) \mathcal{U}_A^{\dagger} (z^+, -\infty^+, \boldsymbol{z} ) \right) \bigg]^{bc} \right \} \Psi_j^{(-)} (z^+, \boldsymbol{z}) + \mathcal{O} (g^2 R^2) \; ,
    \label{Eq:SignatureProjectionsQuark}
\end{gather}
\end{widetext}
which evolves according to
\begin{gather}
    \frac{\partial R_{Q,i}^{(+)} (\boldsymbol{z}_1)}{ \partial \eta} = C_F \mathcal{K}_{\rm RP} (\boldsymbol{z}_{1}, \boldsymbol{z}_{2}) \otimes_{\boldsymbol{z}_2} R^{(+)}_{Q,i} (\boldsymbol{z}_2) + \mathcal{O} (g^4 R^2) \; .
    \label{Eq:FinalReggeization}
\end{gather}
The last result identifies $R^{(+)}_{Q, i} (\boldsymbol{z})$ as the RQI operator and demonstrates that its rapidity evolution is governed solely by the Regge-pole contribution given in eq.~\eqref{Eq:KRegge}. Because corrections to eq.~\eqref{Eq:KRegge}, due to operator mixing, are of order $g^4$, its Regge-pole behaviour is expected to break only at $g^6$, i.e. NNLLA (see~\cite{Altinoluk:2026} for more details). The RQI operator admits a clear Wilson-line interpretation: it corresponds to a quark background-field insertion dressed by the logarithm of the product of two semi-infinite Wilson lines defined on a complementary light-cone support. We emphasize that this degree of freedom is universal and can be extracted, for example, from a generalization of the operator in eq.~(\ref{Eq:AlmostReggeizedQuarkInter}) involving an adjoint Wilson line extending from $z^+$ to $-\infty^+$, corresponding to the gluon-initiated channel. This is demonstrated in the accompanying paper~\cite{Altinoluk:2026}. 

The structure and evolution equation of the negative-signature projection are presented in the accompanying paper~\cite{Altinoluk:2026}, where we also derive the generalized massive kernel and establish its equivalence to the momentum-space formulation of the massive-quark Regge trajectory. For completeness, we report the generalized kernel here:
\begin{gather}
    \mathcal{K}_{\rm RP-mass} (\boldsymbol{z}_{1}, \boldsymbol{z}_{2},m) = \frac{a_s  }{(\boldsymbol{z}_{12}^2)^{d-1}} \frac{2^{1-d/2}}{ \displaystyle \Gamma \left( d/2 \right)}  \bigg[ (m |\boldsymbol{z}_{12}|)^{d/2} \nonumber \\ \times K_{d/2} (m |\boldsymbol{z}_{12}|) + i m \slashed{z}_{12 \perp} (m |\boldsymbol{z}_{12}|)^{d/2-1} K_{d/2-1} (m |\boldsymbol{z}_{12}|) \bigg]  ,
\end{gather}
which reduces to $\mathcal{K}_{\rm RP}$ of eq.~(\ref{Eq:KRegge}) in the massless limit. \\

The essence of the present construction is encoded in
eqs.~(\ref{Eq:SignatureProjectionsQuark})
and~(\ref{Eq:FinalReggeization}), which establish, for the first time,
a Wilson-line realization of quark Reggeization in QCD. \\

{\it Conclusions and outlook}---In this Letter, we established a Wilson-line description of quark
Reggeization within the next-to-eikonal shockwave formalism. Starting from
an identity-changing operator constructed from semi-infinite Wilson lines,
we derived its nonlinear rapidity evolution and identified its
positive-signature projection as the Reggeized-quark interpolating operator.
Its dilute evolution reproduces the one-loop quark trajectory. At finite
$N_c$, the signature projection is essential for removing
Regge-pole-breaking contaminations that first enter at NLLA and isolating
Regge-pole evolution, whereas in the planar limit Reggeization emerges
already for the non-signaturized operator. These results extend the
correspondence between Wilson-line evolution and parton Reggeization to
amplitudes carrying $t$-channel quark quantum numbers. More broadly, these results provide a first step toward a unified description of high-energy QCD amplitudes from Wilson lines with arbitrary $t$-channel quantum numbers.

\begin{figure}[t]
    \centering
    \includegraphics[width=0.25\linewidth]{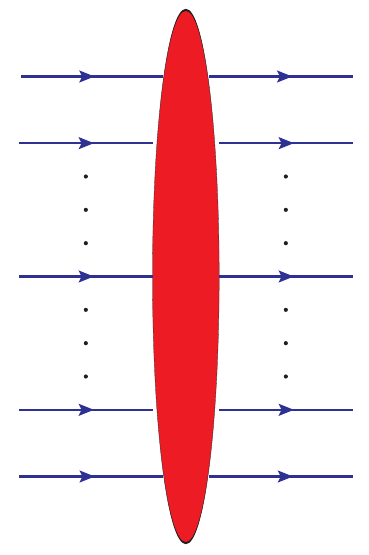} \hspace{1.6 cm}
    \includegraphics[width=0.25\linewidth]{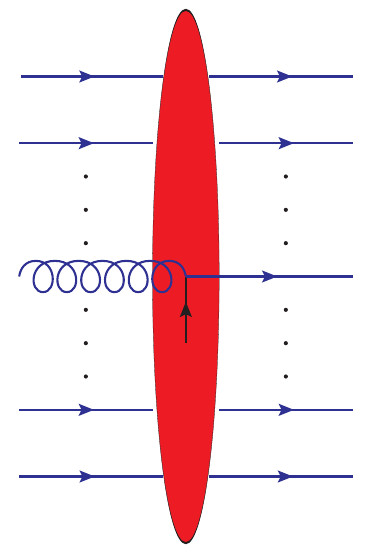} \\
    (a) \hspace{3.3 cm} (b)
    \caption{(a) A projectile built from an arbitrary number of Wilson lines in fundamental representation, whose evolution is described by the Balitsky-JIMWLK equation. (b) A projectile in which one constituent (a Wilson line in adjoint representation) changes its nature via subeikonal interactions with the target quark field (to a Wilson line in fundamental representation).}
    \label{figNeikBalitHiera}
\end{figure}

Several directions naturally emerge from the present work. The immediate extension is the computation of the two-loop massless~\cite{Bogdan:2002sr} and massive quark Regge trajectories. Since two-loop evolution of considerably complicated Wilson-line operators has been derived~\cite{Balitsky:2013fea,Kovner:2013ona,Grabovsky:2013mba} (see also~\cite{Brunello:2025rhh} for progress at three loops), an all-order derivation of quark Reggeization at NLLA now appears within reach.

Finally, the most intriguing direction concerns the role of Regge cuts beyond the accuracy where Regge-pole dominance holds. In the gluon sector, cut contributions can be systematically reconstructed from iterative solutions of the LO Balitsky/JIMWLK evolution~\cite{Caron-Huot:2013fea,Caron-Huot:2017fxr,Caron-Huot:2017zfo,Caron-Huot:2020grv,Falcioni:2020lvv,Falcioni:2021buo,Falcioni:2021dgr}. This suggests an analogous connection in the quark sector and motivates a next-to-eikonal extension of the Balitsky hierarchy in which a Wilson-line constituent undergoes subeikonal transitions into a quark state through interactions with the target background field (see diagram (b) in fig.~\ref{figNeikBalitHiera}). Such a framework may provide access to the all-order structure of quark exchange and multi-Reggeon dynamics, paving the way toward a unified Wilson-line description beyond the eikonal sector.

\section*{Acknowledgments}

We are grateful to R. Boussarie, F. Cougoulic, V. S. Fadin, G. Falcioni, M. A. Nefedov, A. Papa, L. Szymanowski, R. Venugopalan, and S. Wallon for insightful discussions. We further thank V. S. Fadin, G. Falcioni, M. A. Nefedov, and A. Papa for a careful reading of the manuscript. TA and JF are supported in part by the National Science Centre (Poland) under the research Grant No. 2023/50/E/ST2/00133 (SONATA BIS 13). GB is supported in part by the National Science Centre (Poland) under the research Grant No. 2020/38/E/ST2/00122 (SONATA BIS 10). The work of MF is supported by the ULAM fellowship program of NAWA No. BNI/ULM/2024/1/00065 “Color glass condensate effective theory beyond the eikonal approximation”.

\bibliography{mybib_New}

\end{document}